*atmosphere*

MDPI

*Review*

# A Review of Infrasound and Seismic Observations of Sample Return Capsules since the end of the Apollo Era in Anticipation of the OSIRIS-REx Arrival

Elizabeth A. Silber [1,*], Daniel C. Bowman [2] and Sarah Albert [2]

[1] Geophysics, Sandia National Laboratories, Albuquerque, NM 87123, USA
[2] Multi-Modal Geophysics and Data Analytics, Sandia National Laboratories, Albuquerque, NM 87123, USA; dbowma@sandia.gov (D.C.B.); salber@sandia.gov (S.A.)
* Correspondence: esilbe@sandia.gov

**Abstract:** Advancements in space exploration and sample return technology present a unique opportunity to leverage sample return capsules (SRCs) towards studying atmospheric entry of meteoroids and asteroids. Specifically engineered for the secure transport of valuable extraterrestrial samples from interplanetary space to Earth, SRCs offer unexpected benefits that reach beyond their intended purpose. As SRCs enter the Earth's atmosphere at hypervelocity, they are analogous to naturally occurring meteoroids and thus, for all intents and purposes, can be considered artificial meteors. Furthermore, SRCs are capable of generating shockwaves upon reaching the lower transitional flow regime, and thus can be detected by strategically positioned geophysical instrumentation. NASA's OSIRIS-REx (Origins, Spectral Interpretation, Resource Identification, and Security-Regolith Explorer) SRC is one of only a handful of artificial objects to re-enter the Earth's atmosphere from interplanetary space since the end of the Apollo era and it will provide an unprecedented observational opportunity. This review summarizes past infrasound and seismic observational studies of SRC re-entries since the end of the Apollo era and presents their utility towards the better characterization of meteoroid flight through the atmosphere.

## 1. Background

The importance of sample return space exploration missions extends beyond their primary intended mission to collect extraterrestrial (e.g., asteroidal, cometary, or solar) samples, and return them to Earth [1–3]. Specifically, the return vehicles, or sample return capsules (SRCs), are good analogues for a category of extraterrestrial objects impacting and traversing the Earth's atmosphere [4,5]. This analogous behavior is due to SRCs' re-entry trajectory, hypersonic to subsonic velocities, and overall size that lead to the formation of flow fields and shockwaves [6–10]. This additional relevance and utilization of SRCs becomes apparent if one considers that the stochastic nature of natural objects impacting the Earth's atmosphere does not allow for planned comprehensive multi-modal observational campaigns. Rather, only a part of their interaction with the atmosphere is typically detected and recorded with ground- and/or space-based assets. Thus, SRCs serve as useful analogues for natural objects, especially in the context of studying highly nonlinear physics of shockwaves, and for allowing full-spectrum sensing studies of their interaction with the atmosphere, from the moment they impact the upper thermosphere until they land. Moreover, the role of SRCs extends even further during the 'non-luminous' and 'non-ablational' stage of their flight. They may serve as experimental platforms toward developing better detection capabilities for meteoroids and asteroids.

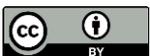

This article has been authored by an employee of National Technology & Engineering Solutions of Sandia, LLC under Contract No. DE-NA0003525 with the U.S. Department of Energy (DOE). The employee owns all right, title and interest in and to the article and is solely responsible for its contents. The United States Government retains and the publisher, by accepting the article for publication, acknowledges that the United States Government retains a non-exclusive, paid-up, irrevocable, world-wide license to publish or reproduce the published form of this article or allow others to do so, for United States Government purposes. The DOE will provide public access to these results of federally sponsored research in accordance with the DOE Public Access Plan https://www.energy.gov/downloads/doe-public-access-plan.

In the remainder of Section 1, we will provide a brief context and elaborate on the importance of SRCs and their parallels to natural meteoric phenomena. In Section 2, we will summarize seismic and infrasound observations of past SRCs, and, in Section 3, outline the upcoming re-entry of NASA's OSIRIS-REx (Origins, Spectral Interpretation, Resource Identification, and Security-Regolith Explorer) SRC and the path forward.

### 1.1. Meteoroids and Asteroids

Earth experiences a continuous influx of extraterrestrial material, encompassing a broad spectrum of particle compositions (ice, rock, metal) and sizes ranging from micron-scale dust grains to macroscopic objects. Upon entry at speeds between 11.2 and 72.5 km/s [5,11], meteoroids and asteroids (>1 m in diameter), herein collectively referred to as meteoroids, undergo complex processes (e.g., sputtering, heating, ablation, fragmentation, hyperthermal and equilibrium chemical reactions with the ambient atmospheric gas) as they encounter denser regions of the atmosphere, e.g., [4,12–18]. The resultant visual phenomenon is called a meteor or a 'shooting star'. Very bright meteors, exceeding the brightness of Venus, are typically produced by objects greater than 10 cm in diameter and are known as fireballs and bolides. These objects also generate shock waves [4,19–21]. In most cases, very small meteoroids completely ablate at altitudes between ~70 and 100 km; however, fireballs and bolides can survive a longer flight through the atmosphere, and in rare cases land in the form of meteorites [5].

The study of meteoroids and asteroids holds practical implications, from better understanding our solar system formation and evolution, to planetary defense. For example, understanding frequency, size distribution, and near-Earth object (NEO) impact hazard is crucial for assessing and mitigating the risks these objects pose to humans, infrastructure, and the biosphere [22–28]. A sobering example of damage that could be inflicted by a NEO is the Chelyabinsk bolide. This event was caused by an ~18 m in diameter asteroid which entered unexpectedly over Russia on February 15, 2013, depositing energy equivalent to 500 kt of trinitrotoluene (TNT) (1 kt TNT = $4.184 \cdot 10^{12}$ J) [23]. For comparison, the yield of the nuclear detonation over Hiroshima in 1945 was ~15 kt of TNT [29]. The bolide produced light and thermal radiation, as well as acoustic and seismic waves. Moreover, because of its shallow (relative to the local horizon) entry, it left an unexpectedly destructive path in its wake, damaging property and causing over 1500 injuries [23,30]. Other events of lesser but still significant energy have happened recently as well. Notable examples include the unannounced energetic bolide event (33 kt) that occurred over Indonesia in 2009 in broad daylight [31], and another unannounced bolide that impacted the Earth's surface in Peru, leaving a crater nearly 14 m across [32–35].

Observational methods to study meteoroids and asteroids include a variety of sensing modalities, including satellites (e.g., geostationary lightning mapper (GLM) and US government sensors), all-sky cameras, radar, infrasound and seismic sensors, and even casual witness reports, e.g., [36–57]. However, none of these sensing modalities provide continuous global coverage nor can they detect meteoroids and asteroids of all sizes. For example, optical observation networks are regionally dense which provides an opportunity to obtain high-fidelity data on a relatively small sample of impacting meteoroids and asteroids [52,58]. However, what is gained on a regional scale is lost on a global scale, given that most of our planet is inaccessible to ground-based camera installations. The problem is compounded by the fact that as the object's size increases, the impact frequency decreases [59,60], and so does a chance to detect it via any given sensing modality in a reasonable time frame and over a designated geographical region. It is also important to emphasize that each sensing modality has inherent limitations which cannot be readily overcome (e.g., all-sky cameras have poor performance during daylight event detections and are rendered unusable under cloudy conditions). Even if ground truth exists, in most cases, impactor characteristics are not well known nor readily derivable, necessitating an implementation of theoretical and numerical models, e.g., [61–63]. Therefore, in order to obtain robust ground truth data and the parameters needed to better characterize an object (e.g., size, velocity, entry angle, composition, energy yield, and orbit), a single event should be recorded by multiple instruments and by more than one sensing modality [52,64,65]. This is rarely achievable, especially considering that asteroid (size range from meters to tens of meters) impacts into our atmosphere are rare and generally happen by chance [23].

Planning multi-instrument observational campaigns aimed at studying and characterizing asteroids in a meter (and greater) size range is neither feasible nor practical because impacts by these objects are sporadic and unannounced (e.g., Chelyabinsk). Even when well-documented by chance, many parameters of interest (e.g., composition, size, porosity, rotation, ablation rate, shock characteristics, hyperthermal chemical processes) and more complex interactions (e.g., ablated and diffusion trail chemical reactions with ambient atmosphere) remain poorly defined, and scientific analyses must rely on assumptions and predictions derived from the theoretical domain [4,54,61,66–69]. Designing an experiment by employing a meter- (or larger) sized object that would serve as an asteroid-analogue and subsequently deploying it from space solely for the purpose of studying NEOs, while technically feasible, is prohibitively costly. Moreover, achieving superorbital speeds would be unattainable. On the other hand, launching small meteoroid-like objects from the lower Earth orbit does not provide a sufficient analogue to NEOs in terms of size or speed. In the 1960s, artificial rocket-borne micrometeoroid experiments using iron and nickel pellets were carried out with the aim to study luminous efficiency [70,71]. However, their sizes were too small to be considered analogous to natural, shock-producing extraterrestrial objects.

This is where SRCs can provide an unprecedented advantage towards studying a well-known meteoric source in all stages of flight (e.g., hypersonic, supersonic, and dark flight). The flight of SRCs is analogous to the entry and descent of meteoroids through our atmosphere and thus can serve as ideal proxies for studying hypersonic shocks produced by natural objects. SRCs are also well-calibrated sources, with known size, mass, and other relevant parameters. Specifically, due to negligible ablation (short of catastrophic disintegration), SRCs preserve their cross section and do not lose mass. This provides an opportunity to refine other detectable (or derivable) parameters that are generally associated with a high degree of uncertainty. Because the typical entry speed of SRCs is in the hypersonic regime that corresponds to the meteoroids on the slower end of the spectrum (~12 km/s), SRCs can be considered artificial meteors. For a reference, the mean speed of natural asteroid entries is 25–30 km/s [72], the lower limit is 11.2 km/s (escape speed from Earth), and the upper limit across bodies of any composition is 72.5 km/s [5,11]. The speeds greater than the uppermost limit are associated with interstellar objects [73], or those coming from outside our Solar System, such as the recent case of the asteroid 'Oumuamua [74].

*1.2. Meteor Generated Shock Waves*

A comprehensive review of meteor-generated shock waves is provided by Silber et al. [4], and, therefore, the discussion on this topic will be kept brief. Meteors generate shock waves through two main mechanisms, generally amplified due to high rate of ablation: hypersonic passage through the atmosphere (or a cylindrical line source) and fragmentation [5,19,75]. Fragmentation can be continuous or discrete (e.g., airburst) [76].

Upon the formation of a shock wave, highly non-linear processes take place in the physical region in the flow-field immediately behind the shock front—this region, called the characteristic or blast radius ($R_0$), delineates the space of maximum energy deposition [19,77,78]. The shock strength can be expressed as a pressure ratio across the shock front, i.e., $\zeta = p/p_0$, where $p$ is the shock pressure and $p_0$ is the ambient pressure [79,80]. In strong shock regime, $\zeta \gg 1$. As it propagates outward, the shock wave loses energy to the surrounding atmosphere, and beyond $10R_0$ decays to a weak shock regime [19,81]. In the weak shock regime, shocks propagate at or near speed of sound ($\zeta \sim 1$) [79,80]. The mathematical expression for the blast radius is $R_0 = (E_0/p_0)^{0.5}$, where $E_0$ is the energy deposition per unit length and $p_0$ is the ambient pressure. If no fragmentation takes place, then $R_0$ can be approximated in terms of meteoroid diameter ($d_m$) and Mach number ($M$) (the ratio of the meteoroid speed and the local speed of sound): $R_0 \sim M d_m$ [19,77,81,82]. It should be noted that there are still some uncertainties pertaining to the relationship between energy deposition and the blast radius, especially in the case of high ablation rates and high velocity. Shock waves produced by meteoroids eventually decay into low frequency acoustic waves ($f < 20$ Hz) or infrasound, which can propagate over great distances due to negligible attenuation ($1/f^2$) [83,84]. Given favorable conditions [85–87], infrasound might be detected by very sensitive ground-based and airborne instruments known as microbarometers [88,89], and the parameters obtained through signal processing used to infer the source energy, and in some cases other attributes (e.g., shock altitude), e.g., [31,43,90]. Infrasound observations of meteor phenomena offer unique advantages. For example, infrasound can propagate over long distances [85], and it is not sensitive to cloud coverage or day/night conditions, thereby facilitating detections where such might not be attainable by other means. Occasionally, meteoroids and asteroids generate seismic waves as well, either through air-to-ground coupling, or a direct impact [33,34].

From an acoustic emissions perspective, the major challenge in characterization of meteoroids and asteroids is the lack of viable source-derived data sets to mine. It should be noted that seismo-acoustic detections of fireballs still rely on the existence of at least some ground truth information, whether that be in the form of high-fidelity observational data from other sensing modalities (e.g., optical) or casual witness reports, or both [43,75]. While great strides over the last couple of decades have been made towards improving detection algorithms, it is profoundly challenging to attribute an observed signal to a specific impulsive source based on infrasound records alone [49,50,61,75]. This is complicated by the fact that meteoroids and asteroids have different intrinsic characteristics (e.g., size, velocity, composition, entry angle, airburst altitude), thus generating signals across a wide range of frequencies, which overlap with many other types of events (e.g., explosions, lightning, volcanoes) [49,50,75].

SRCs, as well-characterized artificial meteors with well-known parameters (e.g., velocity, size, entry angle, spin rate, shape coefficient), can provide important constraints for our models and detection algorithms, and aid in improving sensing techniques [6,7,10]. Arriving on a pre-planned trajectory with ample lead time, SRCs are ideal subjects for dedicated observational campaigns.

## 2. Infrasound and Seismic Observations of Sample Return Capsules

To be considered an analogue for naturally occurring meteoroids and asteroids, an artificial object must enter the Earth's atmosphere at speeds >11 km/s, which is in the velocity range of slow natural meteoroids. This review specifically focuses on objects coming from interplanetary space, and, thus, orbital debris and similar re-entries are excluded from consideration. Also excluded are any space mission re-entries that did not have dedicated and/or published observational campaigns to record infrasound and seismic signatures.

SRCs enter the atmosphere at a very shallow angle, typically 12° or less. The transition from free molecular flow to the lower transitional flow regime and finally the continuum flow regime is commonly accompanied by the initiation of shock waves and the formation of luminous phenomena along its path. Their flight through the atmosphere is initially hypersonic, and then transitions to supersonic, followed by transonic and dark flight regimes before soft landing (Figure 1). This progression in a single re-entry allows for an unparalleled opportunity to investigate different flight stages and assess how acoustic emissions might vary accordingly as well as to conduct comparative studies with other sensing methodologies. Variations in infrasound signal amplitude and frequency content can be used to infer the point of origin of the shock wave, its propagation and attenuation, and atmospheric effects [43,91]. These observations contribute to our understanding of the complex processes occurring as the spacecraft rapidly decelerates and undergoes thermal and mechanical stresses.

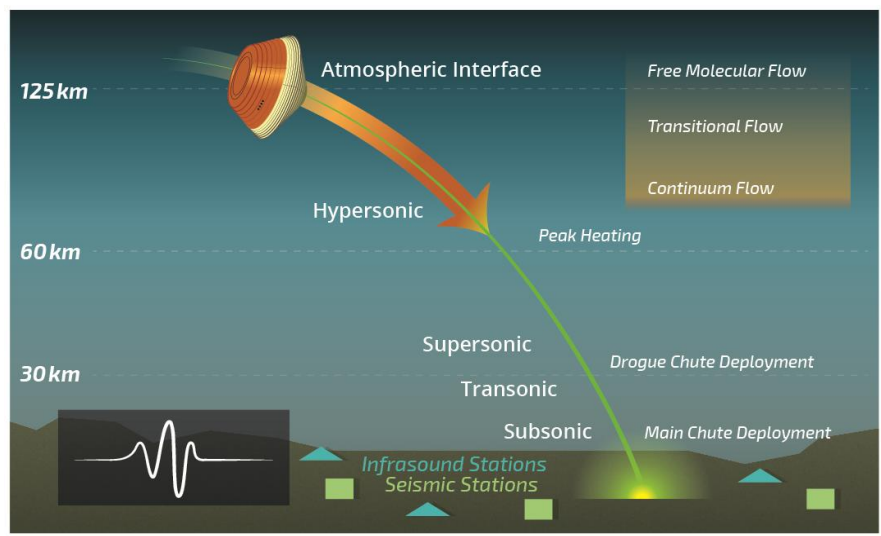

**Figure 1.** Diagram showing the various stages of SRC flight. Diagram not to scale. Infrasound and seismic stations, depending on their location, can detect acoustic signatures generated by the re-entry. The arrow indicates direction of travel.

Since the end of the Apollo era, which ran from 1961 to 1972 [92–94], only four artificial objects re-entering the Earth's atmosphere from interplanetary space were documented seismo-acoustically. All four were SRCs that brought various samples of extraterrestrial material back to Earth; two landed in the USA (Genesis and Stardust) [6,7], and two landed in Australia (Hayabusa 1 and Hayabusa 2) [9,95] (Table 1). The next observational opportunity will present itself on 24 September 2023 with the re-entry of NASA's OSIRIS-REx SRC that will deliver samples of the carbonaceous near-Earth asteroid Bennu [3,96,97]. This section includes a brief review of infrasound and seismic observations of the past SRC re-entries. Table 1 lists the past missions and provides a summary of instruments used to detect the re-entries with microbarometers and seismometers. OSIRIS-REx is also listed, although it is yet to arrive.

**Table 1.** Summary of sample return missions and the number of instruments used towards seismo-acoustic sensing. The landing site abbreviations are: DPG—Dugway Proving Ground, WRC—Woomera Range Complex, UTTR—Utah Test and Training Range. The n/a entries mean that information is not available.

|  | **Genesis** | **Stardust** | **Hayabusa 1** | **Hayabusa 2** | **OSIRIS-REx** |
|---|---|---|---|---|---|
| Launch date | 8 August 2001 | 7 February 1999 | 9 May 2003 | 3 December 2014 | 8 September 2016 |
| Landing date | 8 September 2004 | 15 January 2006 | 13 June 2010 | 5 December 2020 | 24 September 2023 |
| Landing time [UTC] | 15:58 | 10:12 | 14:12 | 05:30 | ~15:00 |
| Landing site | DPG | UTTR | WRC | WRC | UTTR |
| Landing site coordinates [lat, lon] | 40.189° N, −113.213° E | 40.365° N, −113.521° E | 30.955° S, 136.532° E | 30.955° S, 136.532° E | 40.365° N, −113.521° E |
| Entry speed [km/s] | 11 | 12.9 | 12.2 | 12 | 12 |
| Entry angle [°] | 8 | 8.2 | 12 | 12 | ~8 |
| Mass [kg] | 225 | 45.8 | 18 | 16 | 46 |
| Diameter [m] | 1.52 | 0.811 | 0.4 | 0.4 | 0.81 |
| Number of seismometers | 0 | 2 | 20 | 7 | n/a |
| Number of infrasound arrays | 1 | 1 | 1 | 7 | n/a |
| Number of single infrasound sensors | 0 | 0 | 2 | 0 | n/a |
| Combined number of infrasound sensors (single + in arrays) | 3 | 4 | 5 | 28 | n/a |

*2.1. Genesis (2004)*

The Genesis sample return mission was a pioneering endeavor, designed to collect samples of particles present in solar wind and bring them to Earth for analysis [2,98]. The primary scientific objective of the mission was to advance our knowledge about the early solar nebula through elemental and isotopic analysis of the solar particles ejected from the outer layers of the Sun. The Genesis spacecraft was built by Lockheed Martin Space Systems and launched from Kennedy Space Center on 8 August 2001 aboard a Delta 7326 vehicle [2,99]. This space mission hit two major milestones—it was the first to acquire extraterrestrial samples from beyond the Moon's orbit and it was NASA's first sample retrieval since the end of the Apollo program.

In 2004, the re-entry capsule successfully underwent separation and entered the Earth's atmosphere, experiencing a deceleration of 27 g. However, the deployment of the drogue parachute, which serves the purpose of initiating the deployment of the main chute, did not occur as intended because the G-switch was oriented backwards during assembly [100,101]. Consequently, the capsule impacted the ground at an approximate velocity of 311 km/h (86.4 m/s) at 15:58 UTC on 8 September 2004. This error effectively turned Genesis into an artificial meteor along its entire path (and also a meteorite after it hit the surface). NASA's published peak heating point was expected to occur at 15:53:46.25 UTC, at approximately 60.323 km altitude, and at 41.2464° N, −115.7905° E [7]. This was also the point of maximum energy deposition.

A dedicated observational campaign, led by the Los Alamos National Laboratory (LANL) group, consisted of deploying a three-element ground-based infrasound array in Wendover, Nevada (40.7154° N, −114.0357° E) (Figure 2). Simple infrasound array configurations include three or four elements (sensors) in a triangular formation, as shown in Figure 3. Three-element arrays do not have a center element. The reason microbarometers are set up in an array is to determine the wavefront velocity and direction of arrival. Outer (*D*) and inner (*d*) distances between the elements are generally optimized for the predicted wavelength of a given source [102].

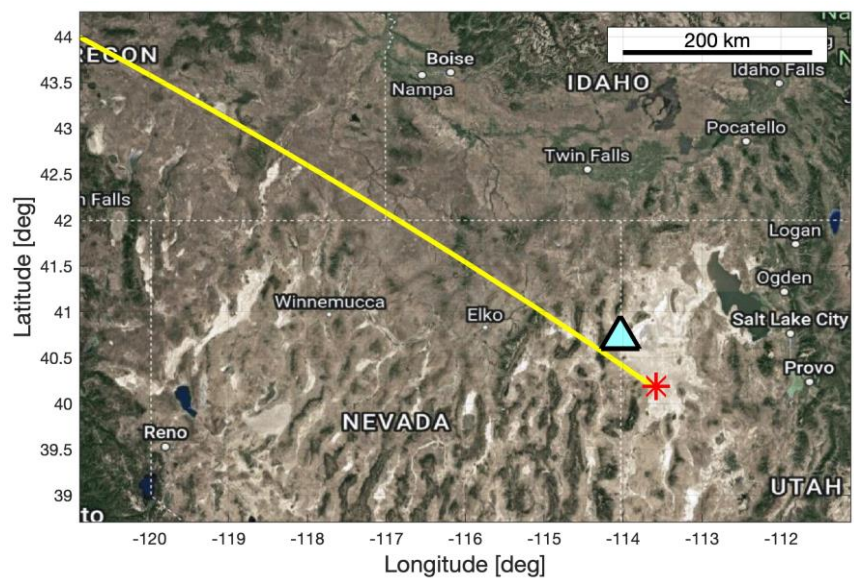

**Figure 2.** Genesis ground track. The location of the infrasound array is shown with the triangle. The landing site is indicated with an asterisk. Map data © 2023 Google.

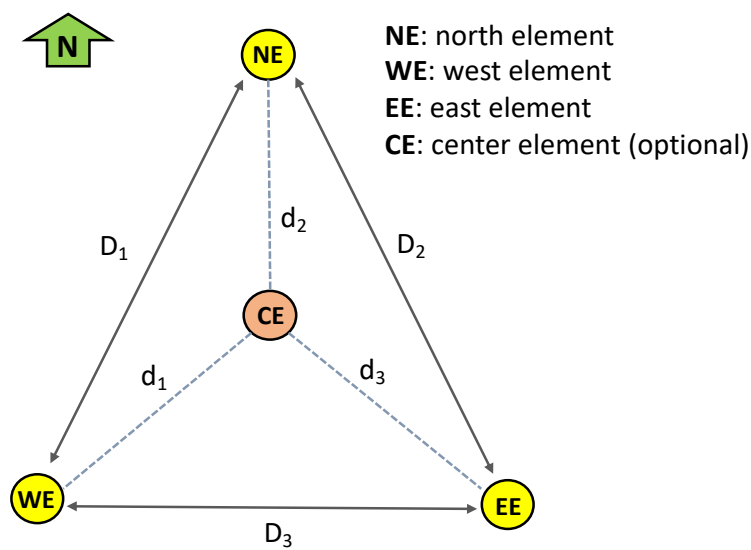

**Figure 3.** Diagram showing a typical triangular configuration of an infrasound array. Circles represent locations of sensors. In the most simple configuration, the center element is absent. Outer (*D*) and inner (*d*) distances between the array elements are generally optimized according to the predicted frequencies of a given source.

The infrasound array was situated 26 km from the nominal ground track (Figure 2) and 43 km relative to the source altitude. The instruments included three Chaparral low-frequency capacitance differential microphones (response: 3 dB down nominally at 0.02 and 300 Hz), a Teledyne-Geotech 24-bit digitizer, and a GPS timing unit. Porous soaker hoses (16-m long) were used to reduce local noise. The digital sampling rate was 50 Hz [7].

The team observing the re-entry heard two distinct sonic booms, but another team member, located nearby, heard only one. The subsequent signal analysis indicated that the main hypersonic boom arrival came at approximately 15:57 UTC, with a delay of 2–4 min (the latter value corresponds to the point of peak heating). The signal recorded at Channel 1 is shown in Figure 4; it exhibits the typical N-wave signature, associated with a ballistic shock. ReVelle et al. [7] concluded that the signal source altitude was 35.5–54.4 km. Trace velocity of the signal was 0.583 km/s. The authors noted that they could not utilize the standard zero-crossing technique to measure the dominant signal period because the results were unreliable; instead, thus they obtained the measurement through fast Fourier transform (FFT) of the signal. The signal properties were as follows: maximum amplitude ($A_{max}$) was 3.99 ± 0.16 Pa, peak-to-peak amplitude ($A_{p2p}$) was 7.26 ± 0.32 Pa, dominant signal period (*P*) was 0.45 ± 0.03 s, and dominant signal frequency was 2.25 ± 0.14 Hz. These are also listed in Table 2, alongside the Stardust signal measurements.

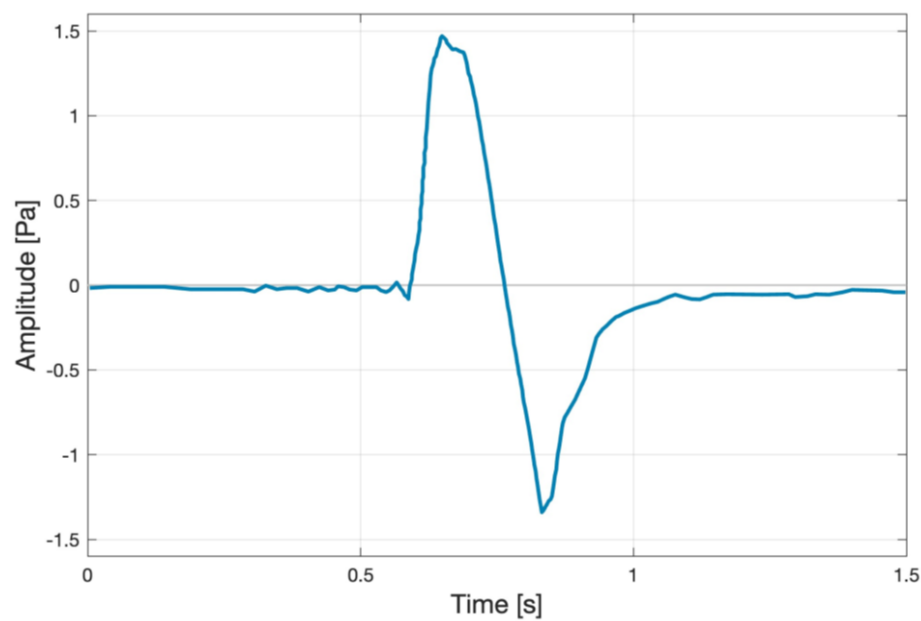

**Figure 4.** Infrasound signal generated by Genesis. The signal arrival time was approximately 15:57 UTC [7]. The N-wave appearance is indicative of the ballistic shock. Figure digitized from actual data collected at site (courtesy of D. ReVelle and P. Brown).

**Table 2.** Infrasound array configurations for observations of Genesis and Stardust. The dimensions are in units of meters. Although the array observing Genesis did not have a center element, ReVelle et al. [7] listed the inner dimensions relative to the nominal center point. These are denoted with the asterisk.

| Distances between array elements | Genesis | Stardust |
| --- | --- | --- |
| $D_1$ | 94.50 | 42.27 |
| $D_2$ | 87.90 | 50.85 |
| $D_3$ | 97.08 | 44.55 |
| $d_1$ | 54.64 * | 26.69 |
| $d_2$ | 55.15 * | 27.35 |
| $d_3$ | 52.30 * | 28.73 |

Due to Genesis' hard landing, the event was expected to yield a seismic signature (surface waves due to the impact or ground motion due to passage of hypersonic shock). Unfortunately, the signal detection algorithms at nearby seismic stations did not consider this event to meet the criteria and no seismic data were saved [7].

*2.2. Stardust (2006)*

The Stardust mission, a part of NASA's Discovery program, was specifically designed to conduct an in-depth study of a comet [1]. The primary scientific objective entailed a close encounter with Comet Wild 2 (also known as 81P/Wild), whereby physical samples from the comet's coma were collected and subsequently transported back to Earth. The spacecraft was launched at 21:04:15 UTC on 7 February 1999, from Cape Canaveral Air Force Station. After a remarkable 7-year journey in space, during which it successfully collected cometary materials, Stardust released the SRC from the main spacecraft on 15 January 2006. Approximately four hours later, the SRC made entry into Earth's atmosphere. While originally intended to be the pioneering robotic mission to bring back samples from a region beyond the Moon's orbit, Stardust ultimately arrived later than the Genesis mission due to its lengthier undertaking. Comparatively, the Stardust SRC had smaller physical dimensions and a reduced mass in contrast to the Genesis SRC. Its diameter accounted for only 53% of the Genesis SRC's diameter, while the mass represented 20% of Genesis SRC. Consequently, the Stardust SRC possessed less energy upon its atmospheric entry phase. The Stardust SRC achieved the highest initial velocity ever recorded for an artificial object entering Earth's atmosphere [6].

The seismo-acoustic observational campaign effort was led by the same LANL group that also led the Genesis deployment [6,7]. The Stardust SRC re-entry on 15 January 2006 was monitored by a four-element ground-based infrasound array accompanied by two co-located seismometers at the airport in Wendover, Nevada (40.7154° N, −114.0357° E) (Figure 5). One seismometer was buried beside the center element, and another one beside the north element. The infrasound array was deployed at a horizontal distance of ~33 km from the nominal trajectory, and it included the same complement of infrasound instruments used for the Genesis observation, with the exception that this array had four Chaparral sensors (Figure 5), and a smaller aperture. The array element separation distances for Genesis and Stardust observations are listed in Table 2 (also see Figure 3). The sampling frequency was 100 Hz [6]. An annotated photo of the center element and the related equipment is shown in Figure 6.

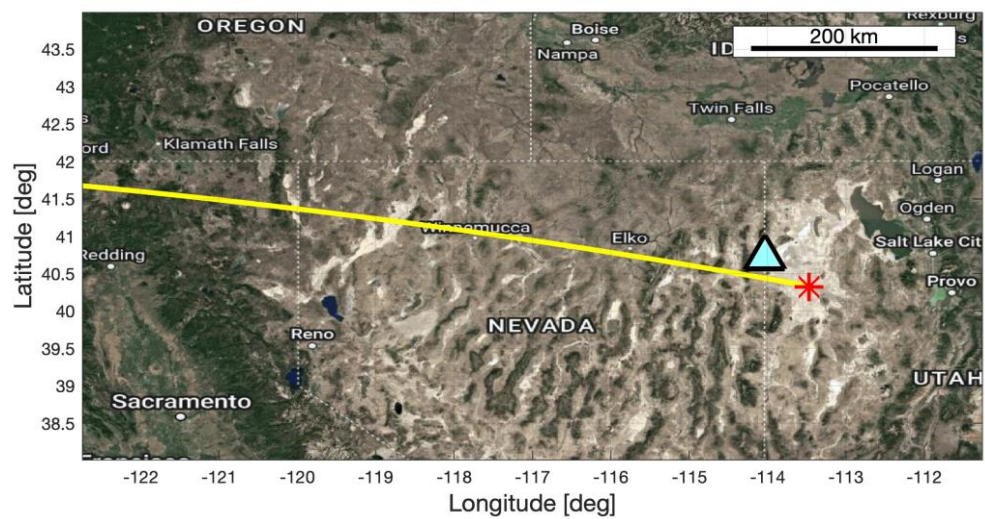

**Figure 5.** Stardust ground track. Triangle shows the location of the array. The landing site is indicated with an asterisk. Map data © 2023 Google.

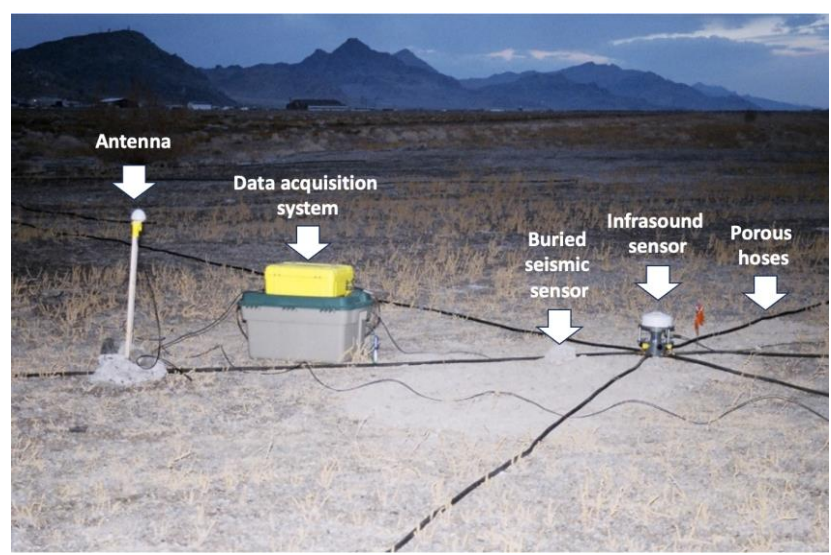

**Figure 6.** Photo of the center element. Various instruments and components are annotated. The seismic sensor is co-located with the center element but not readily visible in the photo because it is buried in the location indicated by the arrow. Porous hoses were used to reduce noise. Photo credit: D. ReVelle and P. Brown.

At the time Stardust signal analysis were performed [6], NASA's peak heating entry time was listed as 09:57:33.42 UTC at an altitude of 61.5 km. ReVelle's entry modeling, however, suggests that the peak heating should have occurred at 57.1 km altitude, which has implications for the post-analysis and determination of the signal source height and travel time.

The entire observing team on site heard a muffled hypersonic boom generated by the Stardust re-entry. Two infrasound signals were detected (Figure 7a), the primary arrival at 10:01:04.2 UTC (Figure 7b), and the secondary arrival (Figure 7c) at 10:01:13.5 UTC (Table 3). The main infrasonic arrival signal parameters were as follows: the peak amplitude was 1.15 ± 0.10 Pa, the peak-to-peak amplitude was 1.80 ± 0.19 Pa, the dominant signal period was 0.20 ± 0.01 s, and the dominant signal frequency was 5.00 ± 0.25 Hz (also see Table 2). The source altitude was estimated at 40 km. Later airwave arrivals in the 1–10 Hz frequency range were also detected and, after detailed analysis, attributed to the ground-to-air coupling [103]. Stardust not only served as a proxy to meteors but also provided the first-ever calibrated measurement of acoustic–seismic coupling efficiency for such an object [103]. The yield estimate, using the AFTAC wave period-based relation, was estimated at $1.206 \times 10^{-2}$ tons of TNT equivalent, and only $6.703 \times 10^{-4}$ tons of TNT equivalent based on the line source (see ReVelle et al. [6] for further details).

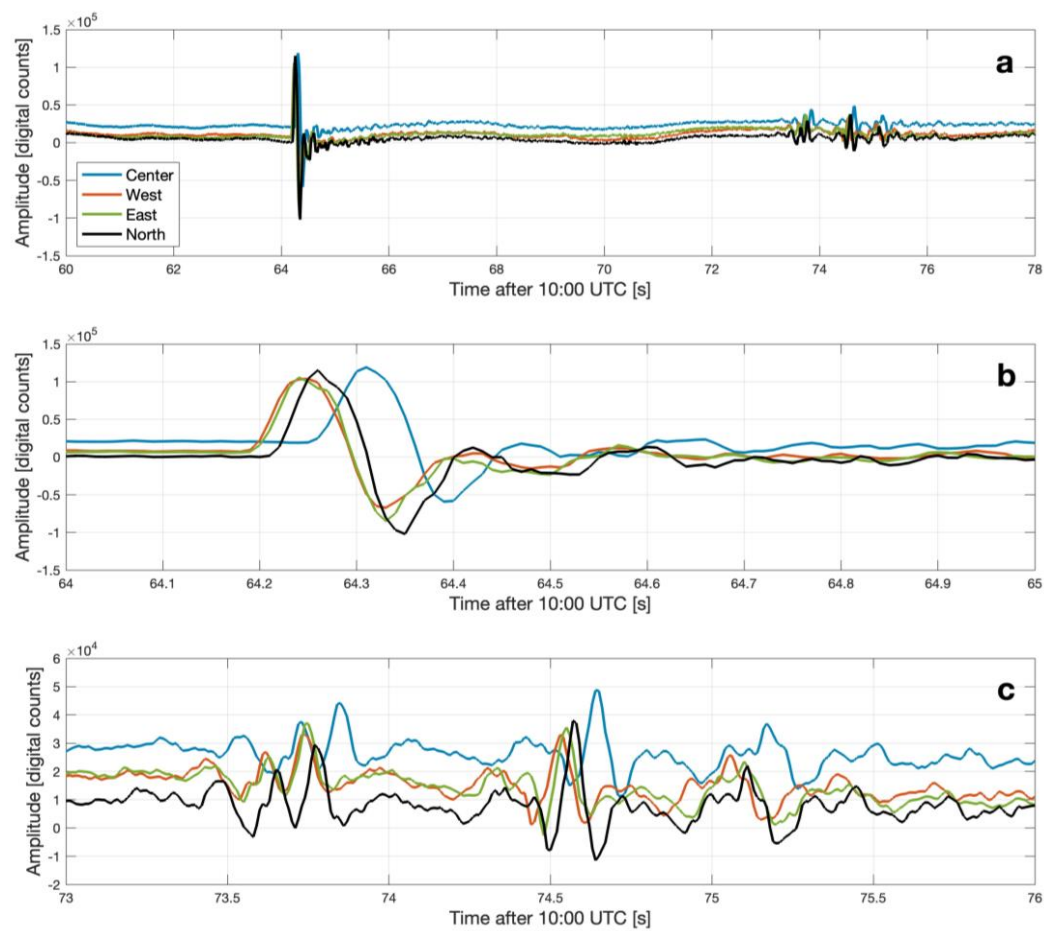

**Figure 7.** (**a**) Infrasound signal arrivals from Stardust. The two lower panels show (**b**) the main and (**c**) later arrivals. Data provided by D. ReVelle and P. Brown (Western University).

**Table 3.** Infrasound signal parameters for Genesis and Stardust (n/a indicates that there is no data).

|  | Genesis | Stardust | |
|---|---|---|---|
|  |  | **Main Arrival** | **Second Arrival** |
| Peak amplitude [Pa] | 3.99 ± 0.16 | 1.15 ± 0.10 | 0.21 ± 0.04 |
| Peak-to-peak amplitude [Pa] | 7.26 ± 0.32 | 1.80 ± 0.19 | n/a |
| Dominant period [s] | 0.45 ± 0.03 | 0.20 ± 0.01 | 0.16 ± 0.01 |
| Dominant frequency [Hz] | 2.25 ± 0.14 | 5.00 ± 0.25 | n/a |

*2.3. Hayabusa 1 (2010)*

The Hayabusa 1 mission was launched on 9 May 2003 by the Japan Aerospace Exploration Agency (JAXA) aboard an M-V vehicle. The mission aimed to study the physical characteristics of the near-Earth asteroid 25143 Itokawa, situated within the Apollo-Amor asteroid class. The primary objectives included sample collection from the asteroid's surface and the return of these samples to Earth.

The projected landing site for the SRC was estimated to cover an area of 20 × 200 km within the confines of the Woomera Prohibited Area, South Australia. The re-entry of the SRC occurred on 13 June 2010 at 13:51 UTC (23:21 local), approximately three hours after it was released from the spacecraft [104]. However, due to prior damage to its chemical propulsion capability, the spacecraft entered along the same trajectory as the SRC. In the three hours following the moment of ejection, a small distance was formed between the spacecraft and the SRC, which allowed the SRC to be tracked independently (Figure 8). As it descended through the atmosphere, the spacecraft experienced a catastrophic breakup at approximately 90 km altitude [104]. Some surviving fragments contributed to the formation of a debris field that dispersed across a defined geographical area. The capsule was recovered in the Australian outback on 14 June 2010.

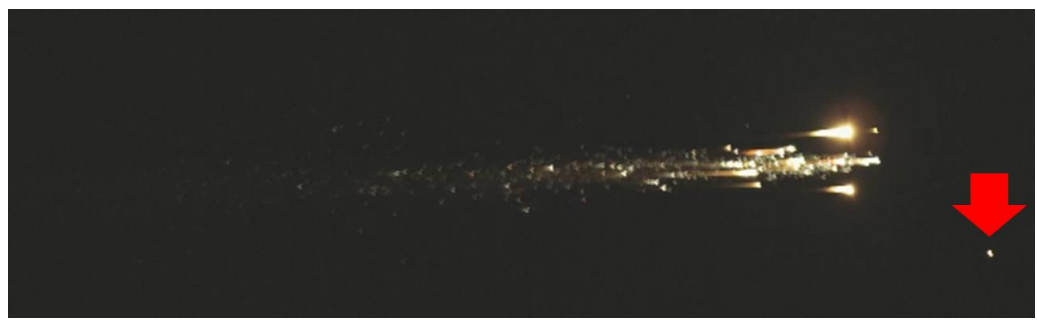

**Figure 8.** A frame from a video that captured the Hayabusa 1 spacecraft debris and the SRC (indicated by the red arrow in the lower right corner). The video was recorded by cameras onboard NASA's DC-8 Airborne Laboratory. Image credit: NASA.

A large complement of instruments was deployed in anticipation of the Hayabusa 1 SRC re-entry [10,95,105]. These included both still and video cameras, infrasound sensors, (audible range) acoustic sensors, and seismometers (Table 1). Figure 9 shows the geographical locations of various deployment sites alongside the Hayabusa 1 SRC ground track, as published by Yamamoto et al. [10]. Optical observation sites are denoted as GOS1, GOS2, GOS3, and GOS4. Infrasound, acoustic (audible range), and seismic sensors were situated at GOS2. A single Chaparral Physics (Model 2) infrasound sensor was installed at GOS2 and another one at GOS2B. A three-element array comprised of Chaparral Model 2.5 sensors was deployed at GOS2A. Soaker hoses were used to reduce the local wind noise. An audible sound recorder was placed at GOS2. Vertical-component-only seismometers (Hakusan, SG820) were installed at GOS2 (5 units), GOS2A (6 units), and GOS2B (7 units), while three-component Sercel (formerly known as Mark Products) L-28-3D seismometers were installed at GOS2 and GOS2A (one unit at each site). Further details are available in Yamamoto et al. [10] and Ishihara et al. [95].

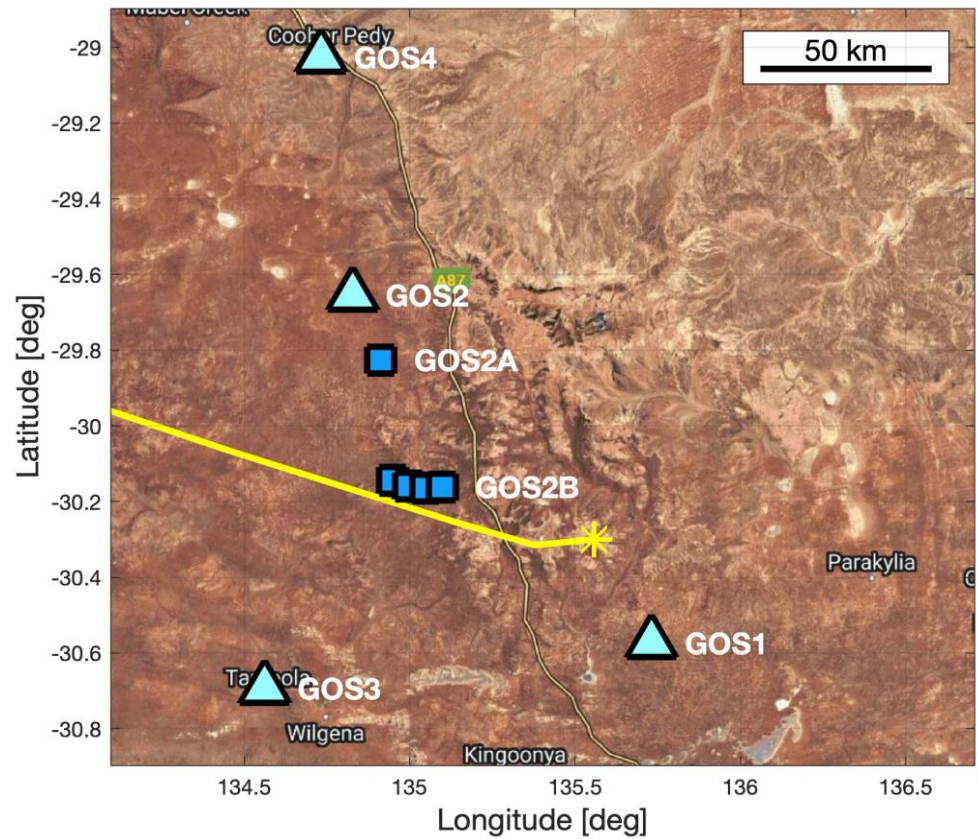

**Figure 9.** Map showing the SRC trajectory, landing site (asterisk), and deployment sites. Triangles denote optical observation points. Squares indicate infrasound/acoustic/seismic instruments were installed at GOS2, GOS2A, and GOS2B. Locations were extracted from Figure 1 published in Yamamoto et al. [10]. Map data © 2023 Google.

The first luminous appearance of Hayabusa was at an altitude of 110 km, and the onset of dark flight occurred at ~36 km [12,30]. Infrasound sensors at all three sites, installed at horizontal distances of 36.9 (GOS2B), 54.9 (GOS2A), and 67.8 (GOS2) km from the trajectory (Figure 9), recorded an N-wave type signature with a peak amplitude of 1.3, 1.0, and 0.7 Pa, respectively [95]. The infrasound signal recorded at the GOS2 site is shown in Figure 10. Seismic signatures were captured as well and determined to be a result of air-to-ground coupling [10,95]. The team at GOS2 heard multiple loud 'bangs' and the team at GOS3 heard audible sounds, as reported by Yamamoto et al. [10] and Ishihara et al. [95]. The GOS2 site where the audible sound was heard was in line with the altitude point of 40.6 km along the SRC trajectory.

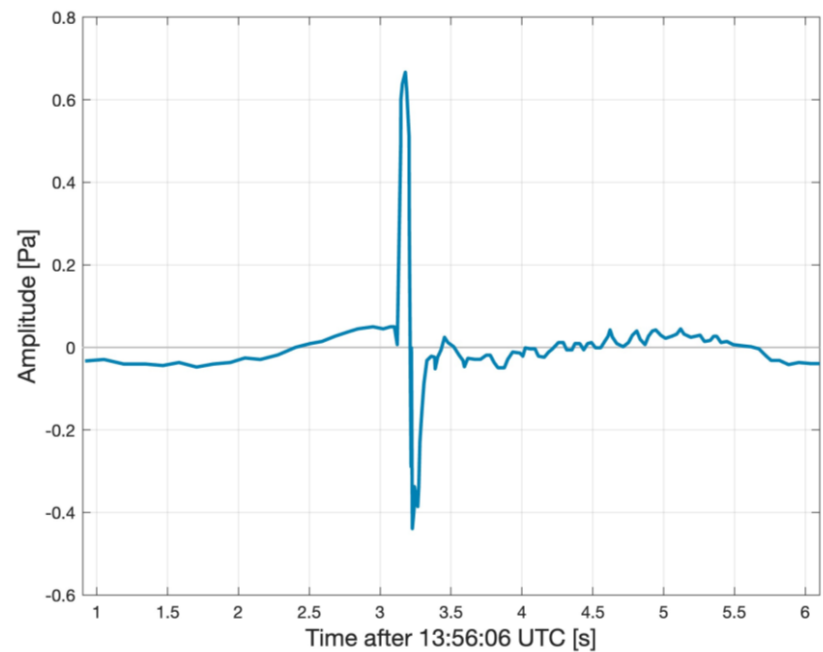

**Figure 10.** Infrasound signal generated by the Hayabusa 1 SRC at GOS2. The waveform is a digitized version of the top panel in Figure 3 from Yamamoto et al. [10].

*2.4. Hayabusa 2 (2020)*

Launched at 04:22:04 UTC on 3 December 2014, the Hayabusa 2 mission aimed to study the near-Earth asteroid 162173 Ryugu, retrieve samples from its surface, and bring those samples to Earth for further analysis. Like its predecessor Hayabusa 1, the Hayabusa 2 SRC was scheduled to land at the Woomera Prohibited Area, South Australia [9,105,106]. The spacecraft released the SRC on 5 December 2020, and with the remaining xenon propellant continued on an extended mission to explore other targets of interest [107].

The arrival of the Hayabusa 2 SRC was recorded by a large complement of instruments (Table 1) with the goal of capturing all aspects of the re-entry [9,106]. Its luminous path lasted 53 s, starting at an altitude of 103 km (17:28:38 UTC), and ending at 39 km (17:29:31 UTC). Beyond the latter point, it assumed dark flight [9]. Compared with the previous three SRC re-entries, this observational campaign utilized a record number of infrasound sensors that were deployed as four-element arrays at seven sites (Figure 11).

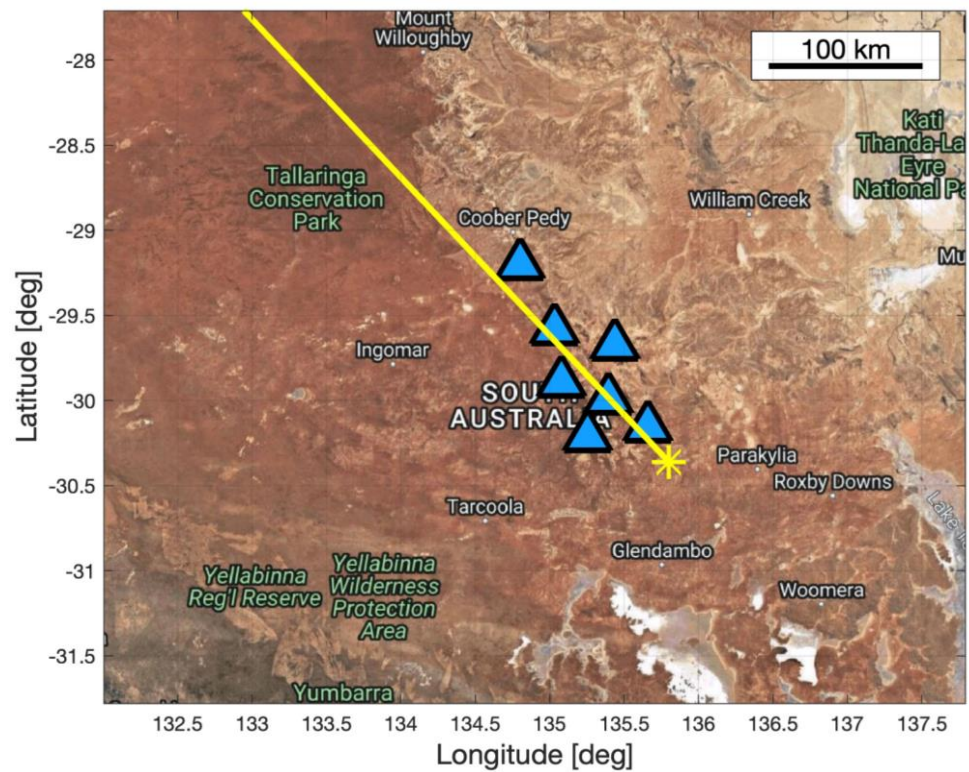

**Figure 11.** Map showing the SRC ground track, landing site (asterisk), and locations of the seven infrasound arrays (triangles). The locations were digitized from Nishikawa et al. [106]. Map data © 2023 Google.

The first signal was detected at 17:31 UTC, with airwaves sweeping the arrays from the southernmost to northernmost one, indicative of shockwave propagation from the trajectory. The N-wave recorded by 27 out of 28 infrasound sensors resembled that of Hayabusa 1. The seismic signatures were captured as well [9,106]. The Hayabusa 2 SRC hypersonic re-entry also generated audible sounds at 18:32:15 UTC that were reminiscent of a mine blast as per observer reports on site as well as casual witnesses in the Coober Pedy township. This region was almost directly underneath the flight path of the SRC [9]. The amount of data collected allowed for a comprehensive analysis, including a 3D trajectory reconstruction based on infrasound detections [106].

## 3. Upcoming Re-Entries and the Path Forward

### 3.1. OSIRIS-REx (2023)

The OSIRIS-REx asteroid sample return mission was launched in 2016 with the aim to collect samples from the near-Earth asteroid Bennu and deliver those samples back to Earth in pristine condition. Bennu was chosen because it is a readily accessible primitive, carbonaceous asteroid, and also one of the most potentially hazardous known near-Earth objects [3,96,97,108]. In terms of design (see Table 1) and planned re-entry, the OSIRIS-REx SRC is identical to the Stardust SRC, and the acoustic signatures are supposed to be similar. The re-entry will consist of several flight phases, including the hypersonic, transonic, and dark flight (see Figure 1). Landing is planned for 24 September 2023 at approximately 14:30 UTC over the region enclosed by an 80 km long and 20 km wide ellipse at UTTR [3,97]. The nominal ground track is shown in Figure 12 (M. Moreau, NASA, personal communication).

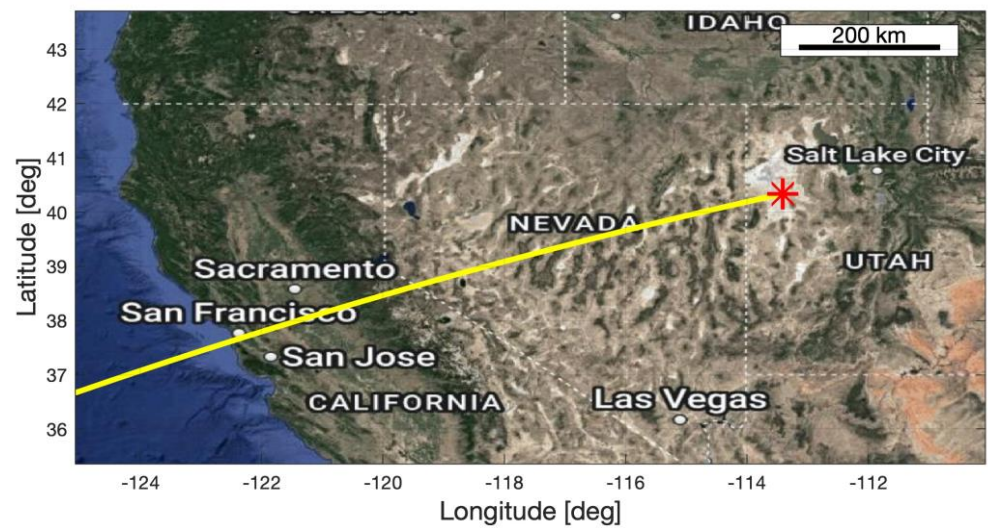

**Figure 12.** Nominal ground track of the OSIRIS-REx re-entry. Red asterisk is the landing site. Map data © 2023 Google.

The atmospheric interface will take place at 14:41:54 UTC (M. Moreau, personal communication) and at an altitude of 125 km. The peak heating is expected to occur at an altitude of approximately 61 km, over the town of Eureka, NV [3,97]. This is also the point of maximum energy deposition, e.g., [109]. The atmospheric passage will last several minutes before the SRC assumes dark flight, at an altitude of 36 km. This is also the point at which the parachute deployment sequence will commence [108]. Unlike previous SRC re-entries, this will be a daylight event (morning hours local time), which will stifle optical observations. Being only the fifth object re-entering from interplanetary space since the end of the Apollo era, OSIRIS-REx will provide an extraordinary opportunity to study meteor phenomena and shockwaves generated at all stages of SRC's hypersonic path, as well as to detect and characterize their signatures via a dense infrasound and seismic sensor network [8].

*3.2. Path Forward*

With ongoing advancements and technological evolution in space exploration, it will be important to capitalize on future events of opportunity. While space missions and other types of exercises might have focused goals and objectives, it is essential to venture beyond their original scope and leverage all possible venues towards studying a broader picture. In this pursuit, the study of shock phenomena assumes paramount significance. The examination of shockwaves on Earth and other planetary bodies, such as Venus [110–112] and Mars [113–115], is essential for elucidating fundamental aspects of impact phenomena, planetary geophysics, and atmospheric dynamics. The utilization of infrasound sensing allows us to study the signatures of shockwaves, thus facilitating a more comprehensive understanding of the underlying physical processes, and characterization of the objects that generate them.

**Author Contributions:** Conceptualization, E.A.S.; writing—original draft preparation, E.A.S.; writing—review and editing, E.A.S., D.C.B., and S.A.; visualization, E.A.S. and D.C.B.; project administration and funding acquisition, E.A.S. and D.C.B. All authors have read and agreed to the published version of the manuscript.

**Funding:** This work was supported by the Laboratory Directed Research and Development program at Sandia National Laboratories, a multimission laboratory managed and operated by National Technology and Engineering Solutions of Sandia, LLC., a wholly owned subsidiary of Honeywell International, Inc., for the U.S. Department of Energy's National Nuclear Security Administration under contract DE-NA-0003525.

**Data Availability Statement:** Data in tables and figures are available in the referenced literature (unless specified otherwise). The Google maps were generated using a modified version of the MATLAB code written by Bar–Yehuda, https://github.com/zoharby/plot_google_map/tree/master (URL accessed on Jun 25, 2023).

**Acknowledgments:** Sandia National Laboratories is a multi-mission laboratory managed and operated by National Technology and Engineering Solutions of Sandia, LLC (NTESS), a wholly owned subsidiary of Honeywell International Inc., for the U.S. Department of Energy's National Nuclear Security Administration (DOE/NNSA) under contract DE-NA0003525. This written work is authored by an employee of NTESS. The employee, not NTESS, owns the right, title, and interest in and to the written work and is responsible for its contents. Any subjective views or opinions that might be expressed in the written work do not necessarily represent the views of the U.S. Government. The publisher acknowledges that the U.S. Government retains a non-exclusive, paid-up, irrevocable, world-wide license to publish or reproduce the published form of this written work or allow others to do so, for U.S. Government purposes. The DOE will provide public access to results of federally sponsored research in accordance with the DOE Public Access Plan. The authors thank M. Moreau (NASA) for providing the preliminary OSIRIS-REx trajectory data, and P. G. Brown (Western University) and late D. O. ReVelle for sharing the Genesis and Stardust trajectory and waveform data. The authors also thank Simon Stähler and the two anonymous reviewers for helpful comments.

**Conflicts of Interest:** The authors declare no conflict of interest.